\documentstyle[sprocl,epsfig]{article}

\def\Journal#1#2#3#4{{#1} {\bf #2}, #3 (#4)}

\def\NPB{{\em Nucl. Phys.} B}

\def\PRD{{\em Phys. Rev.} D}

\def\be{\begin{equation}}
\def\ee{\end{equation}}
\def\bea{\begin{eqnarray}}
\def\eea{\end{eqnarray}}

\begin{document}

\newcommand{\bfk}{\mbox{\bf k}}

\begin{flushright}
INFNCA-TH9701 \\
hep-ph/9701239 \\
January 1997
\end{flushright}
\vspace{1.0truecm}
\renewcommand{\thefootnote}{\fnsymbol{footnote}}
\title{PRODUCTION OF MESON PAIRS INVOLVING
\boldmath{$L\neq 0$} MESONS IN PHOTON-PHOTON COLLISIONS
\footnote{ Talk delivered by F. Murgia at the Diquark III
Workshop, Torino, Italy, October 28-30, 1996.}}

\author{Ladonne Houra-Yaou, Paul Kessler, Joseph Parisi }

\address{ Laboratoire de Physique Corpusculaire, Coll\`ege de France \\
 11, Place Marcelin Berthelot, F-75231 Paris Cedex 05, France}

\author{ Francesco Murgia }

\address{ Istituto Nazionale di Fisica Nucleare, Sezione di Cagliari \\
 Via Ada Negri 18, I-09127 Cagliari, Italy}

\author{ Johan Hansson }

\address{ Department of Physics, Lule\aa{} University of Technology, \\
 S-95187 Lule\aa, Sweden}


\maketitle\abstracts{ We present a formalism for studying the
exclusive production or decay of mesons with any value of the
internal orbital angular momentum $L$. As an application, we
discuss the production of meson pairs (involving tensor and
pseudotensor mesons) in photon-photon collisions.}

\section{Introduction}\label{sec:intro}

In a recent paper \cite{qq} we have presented  a theoretical approach
for the calculation of exclusive processes involving the production or
decay of ($q\,\bar{q}$) mesons with any orbital angular momentum $L$.
Starting from a bound-state model of weakly bound quarks,
a formalism was derived that basically appears as a
natural generalization of the usual perturbative QCD models
for exclusive processes involving hadrons with zero orbital angular
momentum.\cite{br1}
As an interesting application, that formalism was used to study the
production of meson pairs (involving $L\neq 0$ mesons) in photon-photon
collisions. Predictions were given for the corresponding integrated cross
sections in kinematical situations like those of LEP2 or future
B-factories.

Here we will give a shortened presentation of this analysis,
skipping most of the technical details of the approach and
focusing on the qualitative features and on the key steps
of the calculations. The interested reader can find a more
detailed treatment in Ref.~1.

\section{ A generalization of PQCD models for exclusive
processes to the production of \boldmath{$L \neq 0$}
\boldmath{$(q\bar q)$} mesons.}\label{sec:form}

Before presenting our generalized approach, let us
briefly recall the basic ideas underlying perturbative QCD models
for exclusive processes (See Ref. 2 for more details).
These models rely on factorization ideas, applied at the level
of the helicity amplitudes for the physical process considered.
Those amplitudes are then expressed as a convolution among:
{\it i)} a hard-scattering amplitude involving the valence partons
of all participating hadrons, assumed to be, for each hadron,
collinear among themselves and with the parent hadron;
{\it ii)} soft, nonperturbative contributions, described by means of
(distribution) amplitudes for: {\it a)} finding the (collinear)
valence partons in the incoming hadrons; {\it b)} the final
partonic state to form the observed outgoing hadrons.

Notice that: {\it i)} only leading Fock states are considered
(i.e., $|q\bar{q}\rangle$ for mesons, $|qqq\rangle$ for baryons);
{\it ii)} valence partons are taken as massless and in a relative
collinear configuration ($L=0$).
A number of higher-twist effects can in principle modify the
details of the models and play a relevant role at presently
accessible energies.

Let us now discuss our generalized approach;
consider a production process $a(\lambda_a)b(\lambda_b)
\to Q(LSJ\Lambda)\, c(\lambda_c)$, where $Q$ is a $(q\bar{q})$
meson (with quantum numbers $L$, $S$, $J$, and $\Lambda$)
and $a$, $b$, $c$ are any particles (with helicities resp.
$\lambda_a$, $\lambda_b$, $\lambda_c$). Starting from a
bound-state model of weakly bound quarks for $Q$, a prescription
\cite{cahn} relates the (hadronic-level) amplitude
${\cal M}_{\lambda_{a} \lambda_{b} \lambda_{c} }
(E, \Theta)$ to the (partonic-level) amplitude
${\cal T}^{S\,\Lambda_{S}}_{\lambda_{a}\lambda_{b}\lambda_{c}}
(E, \Theta,\bfk , x)$:

\vspace{4pt}

\begin{equation}
 {\cal M}_{\lambda_{a} \lambda_{b} \lambda_{c} }
 (E, \Theta) =  \left(\frac{M_{Q}}{2}\right)^{1/2}
 \int \frac{d^{3}\bfk}{(2\pi)^{3/2}}\,
 \Psi^{\textstyle *} (\bfk)\,\frac{
 {\cal T}^{S\,\Lambda_{S}}_{\lambda_{a}\lambda_{b}\lambda_{c}}
 (E,\Theta,\bfk,x)}
 {\left[(m_{q}^2+\bfk^2)(m_{\bar q}^2+\bfk^2)\right]^{1/4}}\;,
 \label{mabqc}
\end{equation}

\vspace{4pt}

\noindent where: $E$ is the total energy in the c.m. frame of $a$,$b$;
$\Theta$ is the c.m. scattering angle; $M_Q$, $m_q$, $m_{\bar q}$ are
resp. the masses of the $Q$ meson, the quark $q$ and the antiquark
$\bar{q}$; $2\bfk$ is the relative three-momentum of $q$, $\bar{q}$
inside the meson $Q$, in the meson rest-frame; $\Psi(\bfk)$ is
the corresponding meson wavefunction in momentum space.
Notice that in Eq.~(\ref{mabqc}) the $q$, $\bar{q}$ spinors
are already combined to form a total spin state $|S,\Lambda_S\rangle$.
In the spirit of PQCD models, we can now naturally generalize
Eq.~(\ref{mabqc}), defining the $q$, $\bar{q}$ 4-momenta
as follows:

\begin{eqnarray}
 q^{\mu}=xQ^{\mu}+k^{\mu}~~~~~~~~~~~~~~~~~~~~~~~~
 { \bar{q}}^{\mu}=(1-x)Q^{\mu} - k^{\mu}\; ,
 \label{qmu}
\end{eqnarray}
 
\noindent where $Q^{\mu}$ is the $Q$ meson 4-momentum, and
$k^{\mu}$ is the 4-dimensional generalization of $\bfk$.
By introducing the meson distribution amplitude $\Phi_N(x)$
(normalized to unity),
Eq.~(\ref{mabqc}) generalizes to

\vspace{6pt}

\begin{equation}
 {\cal M}_{\lambda_{a} \lambda_{b} \lambda_{c} }
 (E, \Theta) =  \frac{1}{(2M_{Q})^{1/2}}
 \!\!\int \!\!\!\frac{d^{3}\bfk}{(2\pi)^{3/2}}\,
 \Psi^{\textstyle *} (\bfk)\!
 \!\!\int_0^1 \!\!\frac{dx\,\Phi^{\textstyle *}_N(x)}
 {\sqrt{x(1-x)}}\,
 {\cal T}^{S\,\Lambda_{S}}_{\lambda_{a}\lambda_{b}\lambda_{c}}
 (E,\Theta,\bfk,x)\, .
 \label{mabqc2}
\end{equation}

\vspace{6pt}

Using the well-known decompositions: $\Psi(\bfk) = R_{L}(k)\,
Y_{L \Lambda_{L}} ( \theta, \phi)$; and
$|J,\Lambda\rangle = \sum
C^{L\;\;S\;\;J}_{\Lambda_L\,\Lambda_S\,\Lambda}\,|L,\Lambda_L\rangle
|S,\Lambda_S\rangle$ (the $C$'s are the usual Clebsch-Gordan
coefficients), we can now proceed with two crucial steps in our derivation:

\noindent {\bf\it i)} Assuming that in the $a$,$b$ c.m. frame the meson
$Q$ is extreme-relativistic, i.e. $\eta=(M_Q/E) \ll 1$, one can
easily show that the partonic amplitudes ${\cal T}$ become independent
of the azimuthal angle $\phi$; as a consequence, it must be
$\Lambda_L=0$ and $\Lambda=\Lambda_S$.

\noindent {\bf\it ii)} One can notice from Eq.~(\ref{mabqc2})
that the integrand with respect to the absolute value of the relative
$q$,$\bar{q}$ 3-momentum, $k=|\bfk|$, can be expanded in increasing
powers of $k$, the leading term being proportional to $k^L$;
assuming that $R_L(k)$ is sharply peaked towards $k\to 0$, and keeping
only the leading term in the power expansion, one gets from
Eq.~(\ref{mabqc2})

\vspace{6pt}

\begin{eqnarray}
 {\cal M}^{LS J \Lambda}_{\lambda_{a} \lambda_{b} \lambda_{c}}
 (E, \Theta) \!\!& = & \!\!f_{L}^{\textstyle *}\,
 C^{L\;S\;J}_{0\,\Lambda\,\Lambda} \lim_{\beta \to 0}
 \frac{1}{\beta^{L}} \int \frac{ d ( \cos \theta )}{2}\,
 d^L_{0,0}(\theta)\, \nonumber \\
 &&\times \int \frac{\Phi_{N}^{\textstyle *}(x) dx}{\sqrt{x(1-x)}}\,
 {\cal T}^{S\,\Lambda}_{\lambda_{a} \lambda_{b} \lambda_{c}}
 (E, \Theta, \beta, \theta, x)\; ,
 \label{mqcfin}
\end{eqnarray}

\vspace{6pt}

\noindent where we have introduced the dimensionless variable
$\beta=2k/M_Q$, and the normalization constant $f_L$, which is
connected in the usual way to the value at the origin of the
$L$-th derivative of the radial wave function $R_L(r)$.

This equation is the basic result of our formalism.
It can easily be checked that, for $L=0$, Eq.~(\ref{mqcfin})
leads exactly to the same expression as provided by the
usual PQCD models.

\section{An application: meson pair production in photon-photon
collisions}\label{qqp}

We start this section by briefly recalling the physical interest in
studying hadron production in photon-photon collisions.

\subsection{Hadron production in $\gamma\gamma$ collisions}
 \label{subsec:gg}

Photon-photon collisions represent a very useful tool for the study
of hadron production. Basically, the more attractive feature is the
simple, clean initial state, involving only QED interactions,
which allows one to concentrates on the final, hadronic state.
This way, in fact,
some of the more clean tests for PQCD models were proposed.\cite{br1}
It is well known that exclusive $\gamma\gamma \to$ hadron processes can be
studied in $e^+e^-$ colliders. Let us briefly recall some basic
properties of $\gamma\gamma$ processes in this context.

First of all, compare the one-photon annihilation (OPA) contribution
to the two-photon radiation (TPR) one. Due to the photon quantum
numbers, OPA allows to investigate C-odd hadronic final states.
OPA processes are of order $\alpha^2$ but, due to the virtual photon
propagator involved, $\sigma_{OPA}(e^-e^+\to X) \sim 1/s$ (where $s$ is
the lab. energy squared).
On the contrary, TPR processes make possible to study C-even final hadronic
states (being in this sense complementary to the OPA contribution).
Moreover, even though they are of order $\alpha^4$, one finds that
$\sigma_{TPR}(e^-e^+\to e^-e^+ X) \sim \ln^2(s/m_e^2)$: as a consequence,
the TPR contribution already dominates over the OPA one at beam
energies of a few GeV.

Whereas for a given beam energy
the $e^-e^+$ kinematics for the OPA process is fixed, the continuous
spectra of the photon beams in the TPR process allows simultaneous
measurements at different $\gamma\gamma$ invariant masses.
Notice however that this comes to the expenses of collectable statistics
at a given invariant mass; moreover, due to the typical bremsstrahlung
spectrum ($\sim 1/E_{\gamma}$) of the radiated photons, most of
the TPR processes have low invariant masses.

The photon propagators in the TPR process cause the bulk of
photons to be radiated nearly on mass-shell, at small angles
relative to the beam. This means that $e^-e^+$ colliders
effectively provide two colliding beams of quasi-real photons
with luminosities comparable to those of the collider itself.

\subsection{The $\gamma\gamma \to M\bar{M}$ process}
\label{subsec:mm}

Eq.~(\ref{mqcfin}) can be easily generalized to the
process where two ($q\,\bar{q}$) mesons are produced,
$ab\to QQ'$ (herefrom symbols without and with the
``prime'' are pertinent to $Q$ and $Q'$ mesons respectively).
One needs only to substitute particle $c$ in the original derivation
with the meson $Q'$ and repeat the same steps described in the previous
section for the $Q$ meson. We present the corresponding result
directly in the case where the initial particles are two real
photons:

\begin{eqnarray}
 &\!&{\cal M}^{LS J \Lambda,~L'S'J'\Lambda'}
 _{\lambda_{\gamma} \lambda'_{\gamma}}
 (E, \Theta) =
 f_L^{\textstyle *}~f_{L^\prime}^{\prime \textstyle{*}}
 C^{L\;S\;J}_{0\,\Lambda\,\Lambda}
 C^{L'\;S'\;J'}_{0\,\Lambda'\,\Lambda'} \nonumber \\
 &\;\;\;\times&
 \lim_{\beta,~\beta' \to 0}\frac{1}{\beta^{L} \beta^{\prime L'}}
 \int \frac{d ( \cos \theta )}{2} d^L_{0,0}(\theta)
 \int \frac{ d (\cos \theta')}{2} d^{L'}_{0,0}(\theta')
 \nonumber \\ &\;\;\;\times&
 \int \frac{\Phi_{N}^{\textstyle *}(x) dx}{\sqrt{x(1-x)}}
 \int \frac{\Phi_{N}^{\prime\textstyle *}(x') dx'}{\sqrt{x'(1-x')}}\,
 {\cal T}^{S\,\Lambda,\,S'\,\Lambda'}
 _{\lambda_{\gamma} \lambda'_{\gamma}}
 (E, \Theta, \beta, \beta', \theta, \theta', x , x')\; .
 \nonumber \\
 \label{mqq}
\end{eqnarray}

\vspace{6pt}

We can now apply our model to the case of the production,
in photon-photon collisions, of meson pairs involving tensor
and pseudotensor mesons. In order to get numerical results,
we need three basic ingredients of Eq.~(\ref{mqq}):
{\it i)} the hard scattering amplitudes
${\cal T}^{S\,\Lambda,\,S'\,\Lambda'}
_{\lambda_{\gamma} \lambda'_{\gamma}}$;
{\it ii)} the normalization constants $f_L$;
{\it iii)} the meson distribution amplitudes $\Phi_N(x)$.
Let us briefly describe how these quantities are fixed in our approach.

\subsection{Evaluation of partonic amplitudes}\label{subsec:part}

Since the partonic amplitudes are of course independent of
$L$ (the valence constituents of each meson are in a collinear
configuration), we can actually divide our strategy in three steps:
{\it i)} Take the results for the partonic amplitudes as from the
usual PQCD models for $L=0$ mesons; these amplitudes have been
evaluated originally by Brodsky and Lepage,\cite{br2} and are given as
functions of (among other variables) $x$ and $x'$, the fraction
of the meson momentum carried by quark $q$ ($q'$) inside the meson
$Q$ ($Q'$). {\it ii)} Make the substitutions:
$x\to \tilde x = x + (\beta/2)\cos\theta$,
$x'\to \tilde x' = x'- (\beta'/2)\cos\theta'$; it is not difficult
to convince oneself that this corresponds exactly to
the generalized procedure exposed in the previous section, in the limit
$M/E \ll 1$ (here $M$ indicates the $Q$ or $Q'$ mass).
{\it iii)} Finally, perform a series expansion of the resulting
amplitudes in powers of $\beta$, $\beta'$, keeping only the
physically relevant terms (i.e., those in $\beta^L\beta'^{L'}$).

Even so, the expressions of those amplitudes
are quite involved and resort to
numerical integration is required for performing
the convolution integrals.

\subsection{Normalized meson distribution amplitudes, $\Phi_N(x)$}
\label{subsec:phi}

In order to check the dependence of our results on the meson
distribution amplitudes, we have considered two indicative
choices: {\it i)} the so-called nonrelativistic DA,
$\Phi_N(x)=\delta(x-1/2)$. It  leads to very simple
convolution integrals, and in this case we can perform
analytical calculations even for the resulting differential cross sections.
{\it ii)} A generalization of the so-called asymptotic distribution
amplitude,\cite{br1} that is $\Phi_N(x) = N_L x^{L+1}(1-x)^{L+1}$, where
$N_L$ is a factor ensuring the required  normalization to unity.
In particular, for $L=1$, $L=2$ mesons we have respectively
$\Phi_N(x)=30x^2(1-x)^2$, $\Phi_N(x)=140x^3(1-x)^3$.
Since we have considered also the production of hybrid meson
pairs (that is, pairs made of a pion plus a pseudotensor mesons),
we have also to choose a DA for the pion; as an example,
in the following we always use the well-known Chernyak-Zhitnitsky
DA,\cite{cz} $\Phi_{N,\pi}^{CZ}(x)=30x(1-x)(2x-1)^2$.

\subsection{Normalization constants $f_L$}
\label{subsec:fl}

In order to make complete numerical predictions for the
processes of interest, we need finally to fix the values
of the normalization constants $f_L$ appearing in Eq.~(\ref{mqq}).
For the pion, we take the experimental value of the leptonic
decay constant, $f_\pi\cong 93$ MeV, and use the relation
$|f_0| = f_\pi/(2\sqrt{3})$.

For tensor and pseudotensor mesons we evaluate,
using the same theoretical approach, the corresponding
two-photon decay widths, $\Gamma(Q\to\gamma\gamma)$.
Making use of the available experimental data for the masses and
the two-photon decay widths of the $f_2$, $a_2$, $f'_2$ and
$\pi_2$ mesons, one gets estimates of the absolute values of the
corresponding $f_{1,2}$ constants. Both the case of
nonrelativistic and generalized asymptotic DA's have been
taken into account.

\section{Results}
\label{sec:res}

Once we have evaluated the helicity amplitudes for the hadronic
process, 
${\cal M}^{LS J \Lambda,~L'S'J'\Lambda'}_{\lambda_\gamma \lambda'_\gamma}
(E,\Theta)$, we can give predictions for physical observables,
such as the differential cross section
with respect to the scattering angle $\Theta$

\begin{eqnarray}
 \frac{d \sigma^{ \gamma \gamma \to QQ'}(E,\Theta)}
 {d ( \cos \Theta)} & = & \frac{\xi}{128 \pi E^{2}}
 \sum_{\lambda_{\gamma} \lambda_{\gamma}',~\Lambda \Lambda'}
 |~ {\cal M}^{LSJ\Lambda,L'S'J' \Lambda'}
 _{\lambda_{\gamma} \lambda_{\gamma}'}(E, \Theta)|^{2}\; ,
 \label{dsdt}
\end{eqnarray}

\noindent
where meson masses have been neglected in the phase-space factor,
and $\xi$=1/2 if $Q$, $Q'$ are identical particles, $\xi$=1
otherwise.

\begin{figure}[t]
\begin{center}
\epsfig{figure=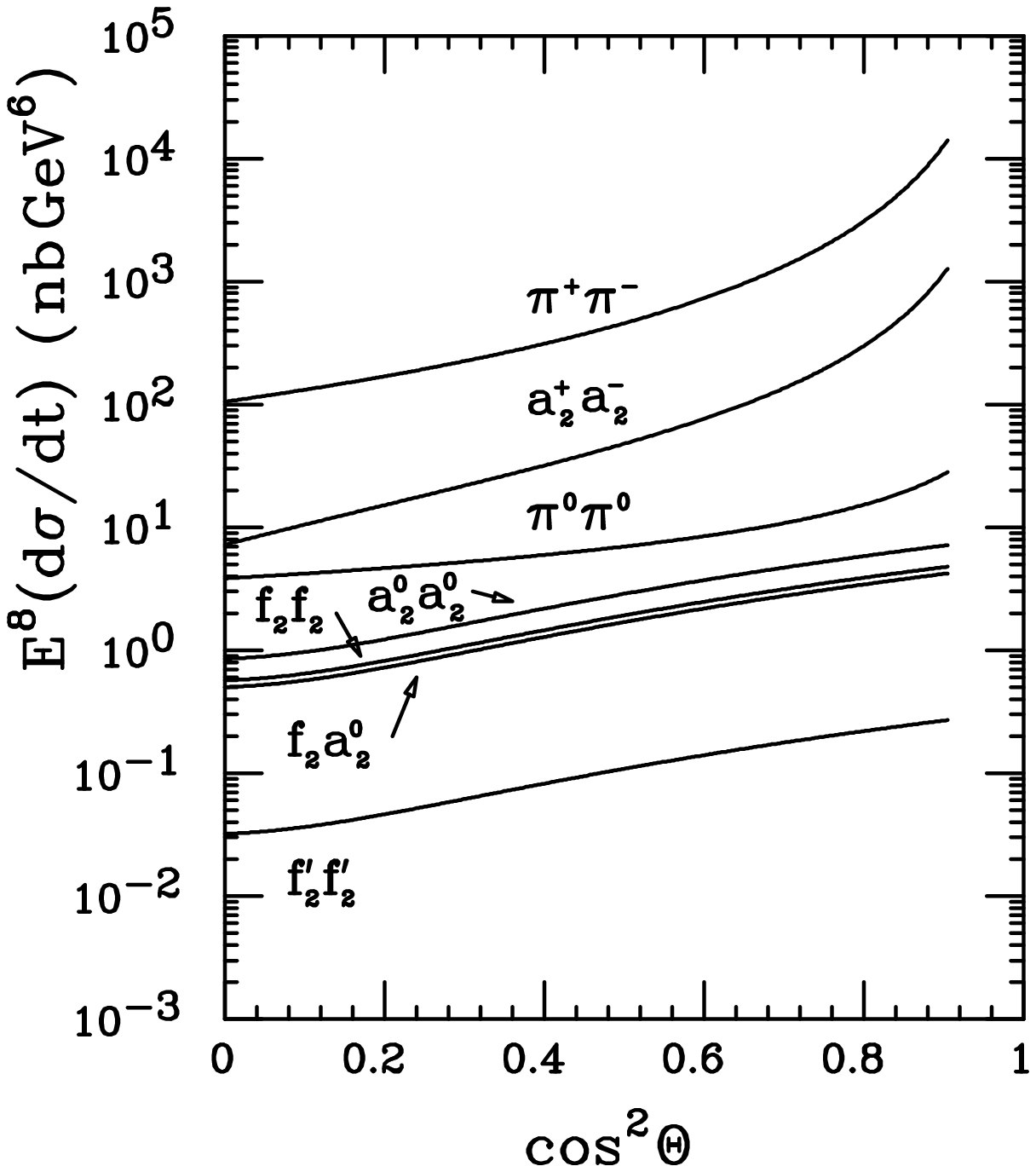,bbllx=80pt,bblly=200pt,bburx=460pt,%
bbury=620pt,width=7cm,height=7cm}
\begin{minipage}[t]{4.0in}
 \baselineskip=8pt
 {\footnotesize {\bf Fig. 1}:
 Differential cross section $E^8\,[d\sigma/dt]$ in
 nb$\times$GeV$^{6}$, as a function of $\cos^2\Theta$,
 for the process $\gamma\gamma\to QQ'$ involving the
 production of tensor-meson pairs; the nonrelativistic
 DA was used for tensor mesons;
 for comparison, analogous curves for
 $\gamma\gamma\to \pi^+\pi^-$ and $\gamma\gamma\to \pi^0\pi^0$,
 using the Chernyak-Zhitnitsky DA are also shown.}
\end{minipage}
\end{center}
\end{figure}

In Fig.s 1-4 we are plotting the scaling differential cross sections
$E^8d\sigma/dt$, which is easily derived from Eq.~(\ref{dsdt})
by noticing that, neglecting masses, $|t|=(E^2/2)(1-\cos\Theta)$.
It is also easy to derive from Eq.~(\ref{dsdt}) the differential
cross section with respect to the transverse momentum $p_T$ of the
outgoing mesons, $d\sigma^{\gamma\gamma\to QQ'}(E,p_T)/dp_T$.
\begin{figure}[t]
\begin{center}
\epsfig{figure=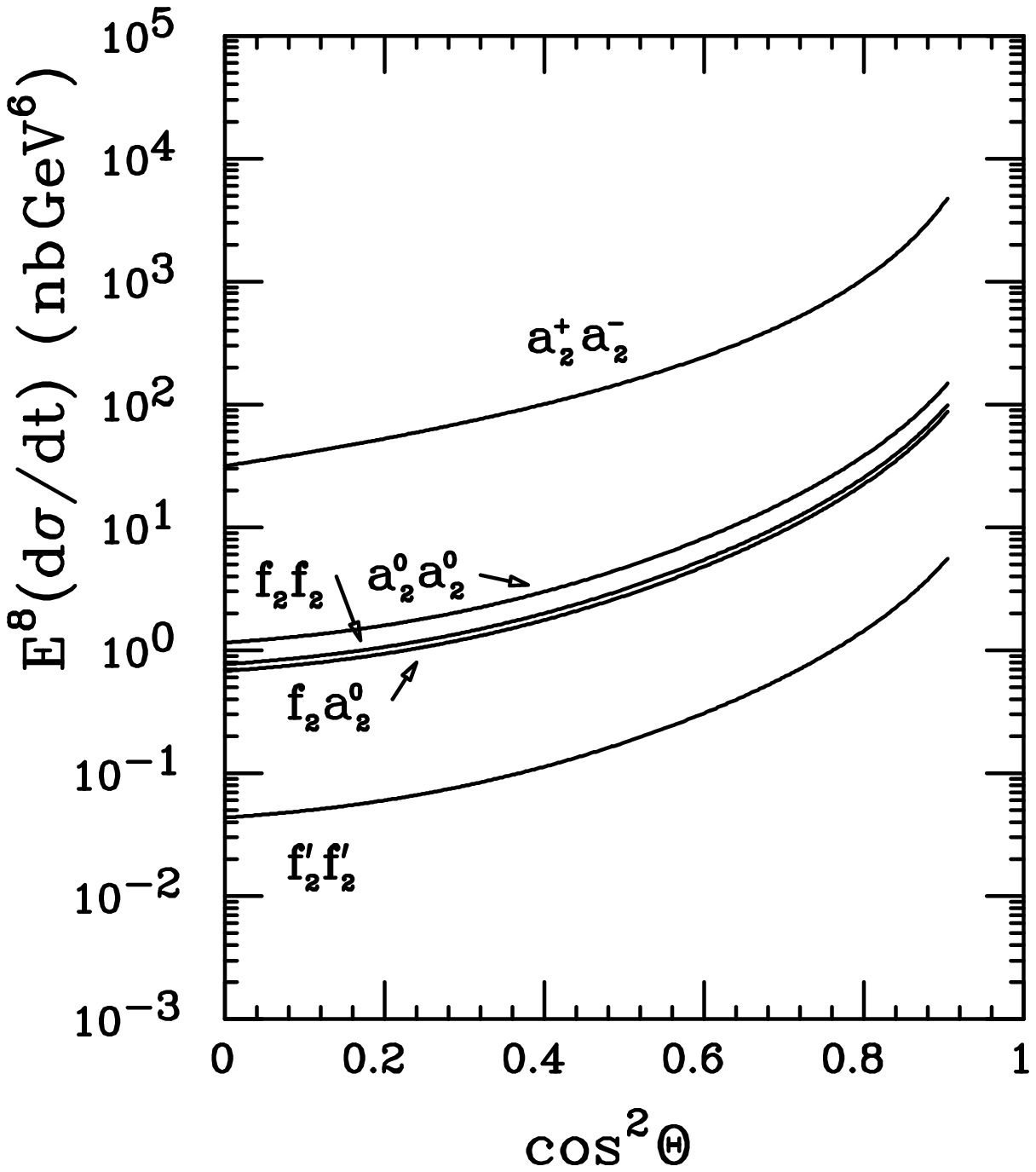,bbllx=80pt,bblly=200pt,bburx=460pt,%
bbury=620pt,width=7cm,height=7cm}
\begin{minipage}[t]{4.0in}
 \baselineskip=8pt
 {\footnotesize {\bf Fig. 2}:
 The same as in Fig. 1, but the generalized asymptotic DA
 is used for tensor mesons.}
\end{minipage}
\end{center}
\end{figure}
The integrated cross section for the overall process $(e^-e^+
\to e^-e^+ QQ')$ can then be obtained, in the equivalent-photon
approximation, by convoluting $d\sigma^{\gamma\gamma\to QQ'}/dp_T$
with the equivalent-photon spectrum of the two photons, and
integrating over $p_T$ from a given minimum value of $p_T$,
$p_T^{min}$. In tables 1-3 we present the results for the
integrated cross section for three cases:
 {\it i)} $\sqrt{s}=200$ GeV (LEP2 energy), $p_T > 1$ GeV~;
{\it ii)} $\sqrt{s}=200$ GeV, $p_T > 2$ GeV~;
{\it iii)} $\sqrt{s}=10$ GeV (energy of a ``B factory''),
$p_T > 1$ GeV.

\section{Conclusions}

We have presented a formalism for the study of exclusive
decay or production processes involving $(q\bar{q})$ mesons
having non-zero orbital angular momentum. Our approach is a
generalization of the usual perturbative QCD models for exclusive
processes involving $L=0$ mesons.

As an application of our model, we have considered in detail
the production of meson pairs (involving tensor, pseudotensor mesons) in
photon-photon collisions.
{}From fig.s 1-4 and tables 1-3 we can argue that the results obtained
do not depend strongly on the distribution amplitude chosen for
tensor and pseudotensor mesons: apart from $\pi_2^0\pi_2^0$
production, the generalized asymptotic DA leads to approximately
equal or slightly (at most by a factor of about 3) higher values,
as compared to the nonrelativistic one.
One can also notice that in general the charged channels give rise
to significantly higher yields than the neutral ones, being thus more
favorable for experimental searches.

Finally, tables 1-3 show that, although the integrated cross sections are
small, there is some hope that the production of charged-meson pairs as here
considered may become measurable with high-energy $e^-e^+$ colliders
of the next generation, provided integrated luminosities
as high as $\approx 10^{40}$ cm$^{-2}$ can be reached.

\begin{figure}[t]
\begin{center}
\epsfig{figure=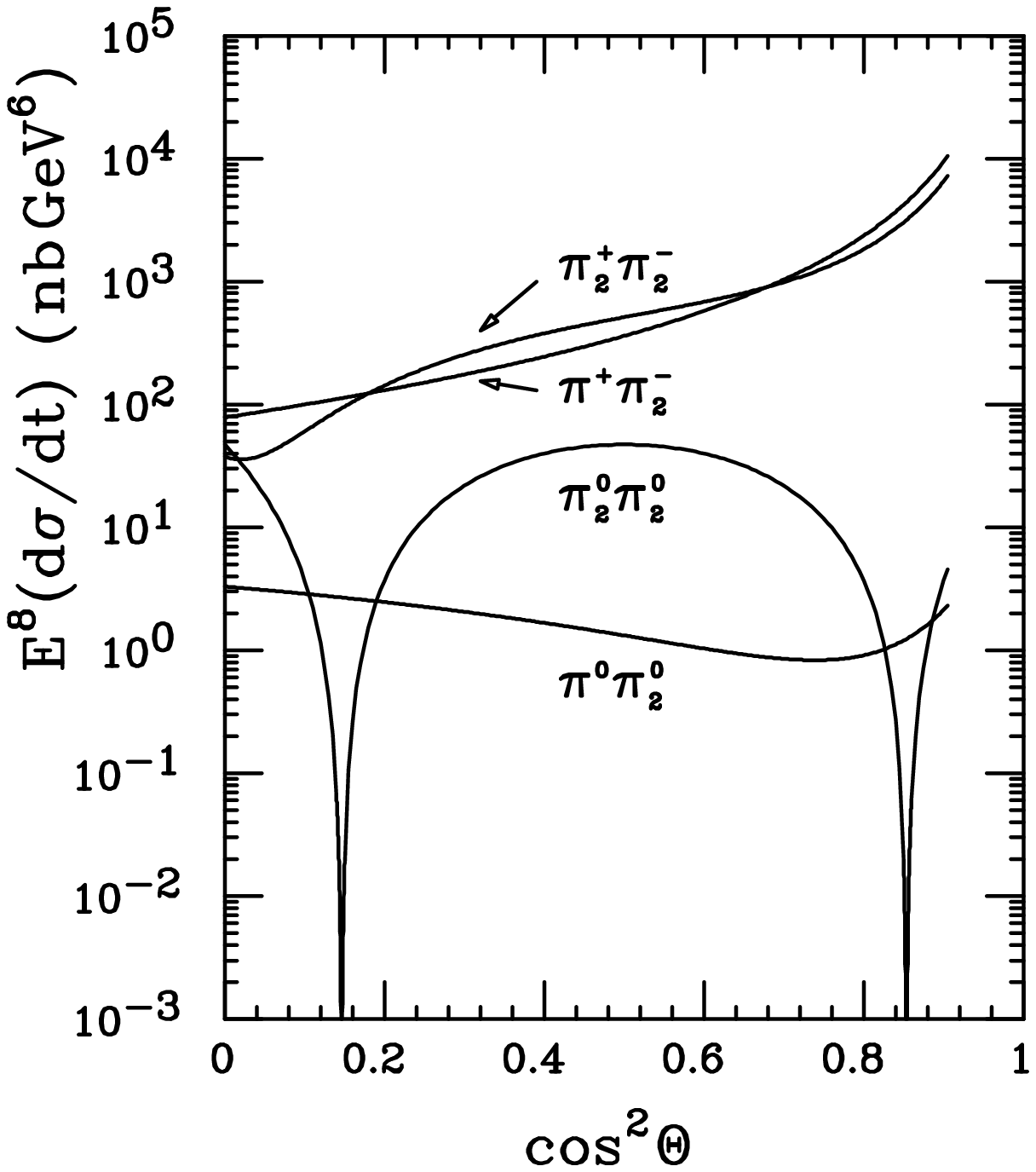,bbllx=80pt,bblly=200pt,bburx=460pt,%
bbury=620pt,width=7cm,height=7cm}
\begin{minipage}[t]{4.0in}
 \baselineskip=8pt
 {\footnotesize {\bf Fig. 3}:
 Differential cross section $E^8\,[d\sigma/dt]$ in
 nb$\times$GeV$^{6}$, as a function of $\cos^2\Theta$,
 for the process $\gamma\gamma\to QQ'$ involving the
 production of pseudotensor-meson and
 hybrid (one pion plus one pseudotensor meson) pairs.
 The Chernyak-Zhitnitsky DA was used
 for pions, while the nonrelativistic DA was used for
 pseudotensor mesons.}
\end{minipage}
\end{center}
\end{figure}

\begin{figure}[t]
\begin{center}
\epsfig{figure=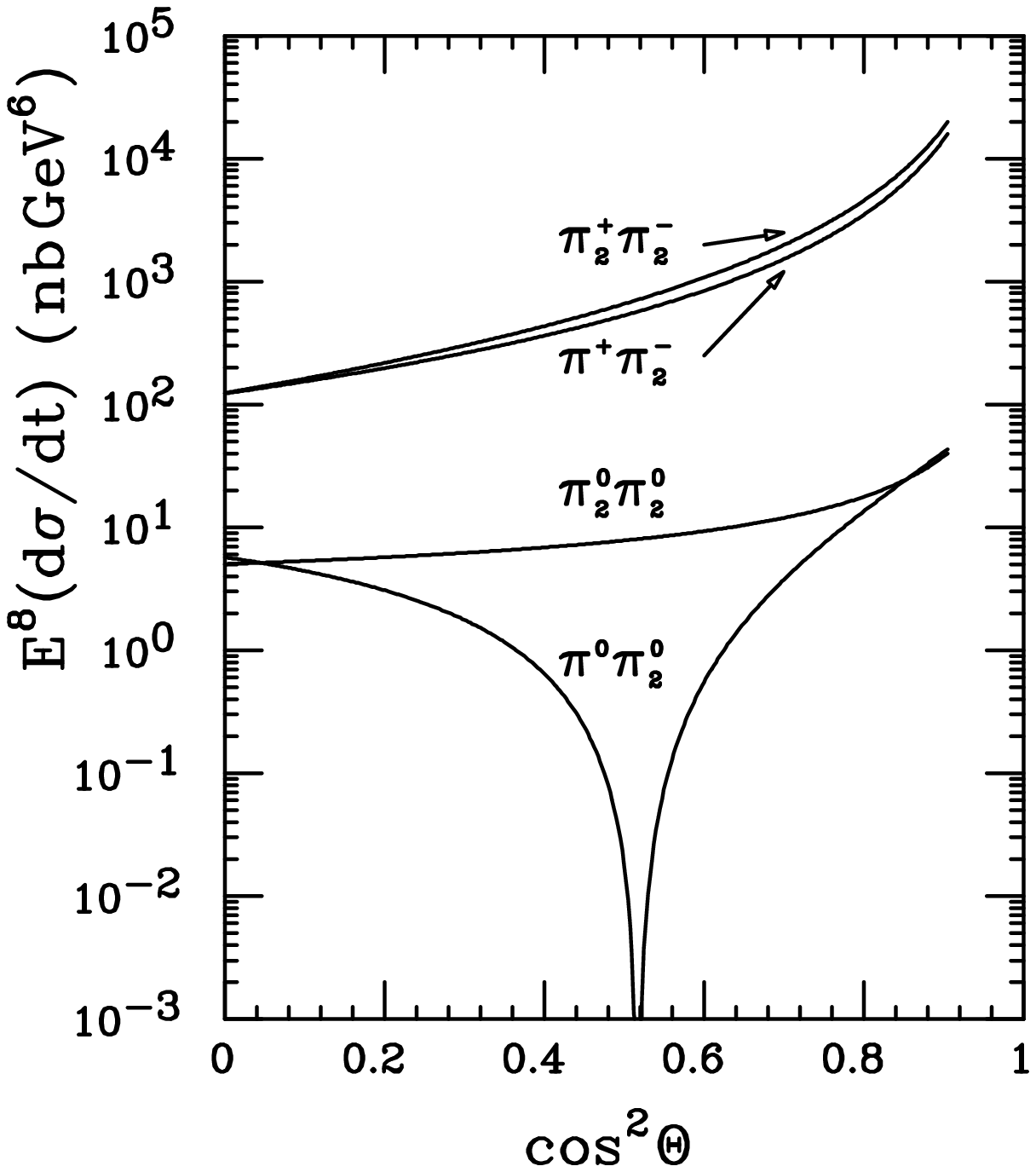,bbllx=80pt,bblly=200pt,bburx=460pt,%
bbury=620pt,width=6.5cm,height=6.5cm}
\begin{minipage}[t]{4.0in}
 \baselineskip=8pt
 {\footnotesize {\bf Fig. 4}:
 The same as in Fig. 3, but the generalized asymptotic DA
 is used for pseudotensor mesons.}
\end{minipage}
\end{center}
\end{figure}

\vspace{-8pt}

\section*{Acknowledgments}
Three of us (J.H., F.M. and J.P.) wish to thank
M.~Anselmino and E.~Predazzi for their kind hospitality
during this very interesting workshop.

This work has been partially supported by the EU program
``Human Capital and Mobility'' under contract CHRX-CT94-0450.

\newpage

\begin{center}
 \begin{minipage}[t]{4.0in}
 \baselineskip=8pt
 {\footnotesize {\bf Table 1}:
 Integrated cross sections (in $10^{-40}$ cm$^2$) of the
 process $ee'\to ee'QQ'$, for $\sqrt{s}=200$ GeV, $p_T > 1$ GeV.}
 \vspace{0.4cm}
 \end{minipage}
 \nopagebreak[4] 
 \small
 \begin{tabular}{crr}
  \hline\hline
  \noalign{\vspace{8pt}}
    $\quad\qquad QQ'\;\qquad$ &
    \multicolumn{2}{c}{$\quad\sigma(ee'\to ee'QQ')
                       \;[10^{-40}$ cm$^2]\quad$} \\
  \noalign{\vspace{4pt}}
  \cline{2-3}
  \noalign{\vspace{4pt}}
    & \multicolumn{1}{c}{$\quad$ NR} & \multicolumn{1}{c}{$\qquad$GASY}  \\
  \noalign{\vspace{4pt}}
  \hline
  \noalign{\vspace{8pt}}
    $f_2\,f_2$         &    35.2   &    49.1 $\qquad\quad$   \\
  \noalign{\vspace{4pt}}
    $a^0_2\,a^0_2$     &    49.7   &    92.6 $\qquad\quad$   \\
  \noalign{\vspace{4pt}}
    $f_2\,a^0_2$       &    31.1   &    57.5 $\qquad\quad$   \\
  \noalign{\vspace{4pt}}
    $f'_2\,f'_2$       &     1.0   &     2.1 $\qquad\quad$   \\
  \noalign{\vspace{4pt}}
    $a^+_2\,a^-_2$     &   494.6   &  1721.8 $\qquad\quad$   \\
  \noalign{\vspace{4pt}}
    $\pi^0_2\,\pi^0_2$ &   236.1   &    37.7 $\qquad\quad$   \\
  \noalign{\vspace{4pt}}
    $\pi^+_2\,\pi^-_2$ &  1387.6   &  2698.2 $\qquad\quad$   \\
  \noalign{\vspace{4pt}}
    $\pi^0\,\pi^0_2$   &   165.8   &   389.6 $\qquad\quad$   \\
  \noalign{\vspace{4pt}}
    $\pi^+\,\pi^-_2$   &  6651.7   & 10476.8 $\qquad\quad$   \\
  \noalign{\vspace{8pt}}
  \hline\hline
 \end{tabular}
\end{center}

\clearpage

\begin{center}
 \begin{minipage}[t]{4.0in}
 \baselineskip=8pt
 {\footnotesize {\bf Table 2}:
 Same as table 1, but assuming $p_T > 2$ GeV.}
 \vspace{0.4cm}
 \end{minipage}
 \nopagebreak[4] 
\small
\begin{tabular}{crr}
  \hline\hline
  \noalign{\vspace{8pt}}
    $\quad\qquad QQ'\;\qquad$ &
    \multicolumn{2}{c}{$\quad\sigma(ee'\to ee'QQ')
                       \;[10^{-40}$ cm$^2]\quad$} \\
  \noalign{\vspace{4pt}}
  \cline{2-3}
  \noalign{\vspace{4pt}}
    & \multicolumn{1}{c}{$\quad$ NR} & \multicolumn{1}{c}{$\qquad$GASY}  \\
  \noalign{\vspace{4pt}}
  \hline
  \noalign{\vspace{8pt}}
    $f_2\,f_2$         &     1.2   &     1.2 $\qquad\quad$   \\
  \noalign{\vspace{4pt}}
    $a^0_2\,a^0_2$     &     2.3   &     3.5 $\qquad\quad$   \\
  \noalign{\vspace{4pt}}
    $f_2\,a^0_2$       &     1.4   &     2.1 $\qquad\quad$   \\
  \noalign{\vspace{4pt}}
    $f'_2\,f'_2$       &     0.1   &     0.1 $\qquad\quad$   \\
  \noalign{\vspace{4pt}}
    $a^+_2\,a^-_2$     &    20.3   &    68.9 $\qquad\quad$   \\
  \noalign{\vspace{4pt}}
    $\pi^0_2\,\pi^0_2$ &    21.6   &     4.1 $\qquad\quad$   \\
  \noalign{\vspace{4pt}}
    $\pi^+_2\,\pi^-_2$ &    92.6   &   164.2 $\qquad\quad$   \\
  \noalign{\vspace{4pt}}
    $\pi^0\,\pi^0_2$   &     6.5   &    13.4 $\qquad\quad$   \\
  \noalign{\vspace{4pt}}
    $\pi^+\,\pi^-_2$   &   189.6   &   298.4 $\qquad\quad$   \\
  \noalign{\vspace{8pt}}
  \hline\hline
 \end{tabular}
\end{center}

\vspace{1.0truecm}

\begin{center}
 \begin{minipage}[t]{4.0in}
 \baselineskip=8pt
 {\footnotesize {\bf Table 3}:
Same as table 1, but assuming: $\sqrt{s}=10$ GeV, $p_T > 1$ GeV.}
 \vspace{0.4cm}
 \end{minipage}
 \nopagebreak[4] 
\small
 \begin{tabular}{crr}
  \hline\hline
  \noalign{\vspace{8pt}}
    $\quad\qquad QQ'\;\qquad$ &
    \multicolumn{2}{c}{$\quad\sigma(ee'\to ee'QQ')
                       \;[10^{-40}$ cm$^2]\quad$} \\
  \noalign{\vspace{4pt}}
  \cline{2-3}
  \noalign{\vspace{4pt}}
    & \multicolumn{1}{c}{$\quad$ NR} & \multicolumn{1}{c}{$\qquad$GASY}  \\
  \noalign{\vspace{4pt}}
  \hline
  \noalign{\vspace{8pt}}
    $f_2\,f_2$         &     2.3   &     2.6 $\qquad\quad$   \\
  \noalign{\vspace{4pt}}
    $a^0_2\,a^0_2$     &     3.6   &     5.9 $\qquad\quad$   \\
  \noalign{\vspace{4pt}}
    $f_2\,a^0_2$       &     2.3   &     3.7 $\qquad\quad$   \\
  \noalign{\vspace{4pt}}
    $f'_2\,f'_2$       &     0.1   &     0.1 $\qquad\quad$   \\
  \noalign{\vspace{4pt}}
    $a^+_2\,a^-_2$     &    33.0   &   113.2 $\qquad\quad$   \\
  \noalign{\vspace{4pt}}
    $\pi^0_2\,\pi^0_2$ &    16.8   &     2.7 $\qquad\quad$   \\
  \noalign{\vspace{4pt}}
    $\pi^+_2\,\pi^-_2$ &    74.0   &   131.8 $\qquad\quad$   \\
  \noalign{\vspace{4pt}}
    $\pi^0\,\pi^0_2$   &    17.1   &    35.9 $\qquad\quad$   \\
  \noalign{\vspace{4pt}}
    $\pi^+\,\pi^-_2$   &   524.4   &   825.2 $\qquad\quad$   \\
  \noalign{\vspace{8pt}}
  \hline\hline
 \end{tabular}
\end{center}

\end{document}